\documentclass[prd,twocolumn,aps,nofootinbib,eqsecnum,notitlepage,nobibnotes,superscriptaddress,preprintnumbers,floatfix]{revtex4-1}

\usepackage{subfig}
\usepackage[a4paper, left=20mm,right=20mm,top=20mm,bottom=20mm]{geometry}

\usepackage{mathtools}
\usepackage{tensor}
\usepackage{amsmath, amsfonts,bm}
\usepackage{derivative}
\usepackage{bm}
\usepackage{amssymb}

\usepackage{suffix}

\usepackage{amsmath, amsfonts,bm}
\usepackage{color}
\usepackage{graphicx}
\usepackage{subfig}
\definecolor{myred}{rgb}{0.9, 0.17, 0.31}

\usepackage[a4paper, left=20mm,right=20mm,top=20mm,bottom=20mm]{geometry}
\usepackage[colorlinks=true,hyperfootnotes=true, linkcolor=myred, citecolor=cyan, urlcolor=magenta]{hyperref}

\newcommand\mathcomma{\,,}
\newcommand\mathperiod{\,.}

\DeclareMathAlphabet{\mathup}{OT1}{\familydefault}{m}{n}

\renewcommand\vec[1]{\bm{#1}}

\newcommand{\be}{\begin{equation}} 
\newcommand{\ee}{\end{equation}}

\begin{document}

\title{Relaxing cosmological constraints on current neutrino masses}

\author{Vitor da Fonseca}
\affiliation{Instituto de Astrof\'isica e Ci\^encias do Espa\c{c}o,\\ 
Faculdade de Ci\^encias da Universidade de Lisboa,  \\ Campo Grande, PT1749-016 
Lisboa, Portugal}
\email{vitor.dafonseca@alunos.fc.ul.pt}
\author{Tiago Barreiro}
\affiliation{Instituto de Astrof\'isica e Ci\^encias do Espa\c{c}o,\\ 
Faculdade de Ci\^encias da Universidade de Lisboa,  \\ Campo Grande, PT1749-016 
Lisboa, Portugal}
\affiliation{ECEO, Universidade Lus\'ofona, \\ Campo Grande, 376,  PT1749-024 Lisboa, Portugal}
\author{Nelson J. Nunes}
\affiliation{Instituto de Astrof\'isica e Ci\^encias do Espa\c{c}o,\\ 
Faculdade de Ci\^encias da Universidade de Lisboa,  \\ Campo Grande, PT1749-016 
Lisboa, Portugal}

\begin{abstract}
We show that a mass-varying neutrino model driven by scalar field dark energy relaxes the existing upper bound on the current neutrino mass to  ${\sum m_\nu < 0.72}$ eV. We extend the standard $\Lambda$ cold dark matter model by introducing two parameters: the rate of change of the scalar field with the number of $e$-folds and the coupling between neutrinos and the field. We investigate how they affect the matter power spectrum, the cosmic microwave background anisotropies and its lensing potential. The model is tested against Planck observations of temperature, polarization, and lensing, combined with baryon acoustic oscillation measurements that constrain the background evolution. The results indicate that small couplings favor a cosmological constant, while larger couplings favor a dynamical dark energy, weakening the upper bound on current neutrino masses.
\end{abstract}

\maketitle
\section{Introduction}\label{sec:intro}

The standard hot big bang model predicts that the Universe is filled with a background of thermal relic neutrinos, called the cosmic neutrino background, with a temperature and density of the order of the cosmic microwave background (CMB) photons \cite{Lesgourgues:2006nd,10.3389/fphy.2017.00070}. Neutrinos are held in thermal equilibrium with the primordial plasma by electroweak interactions until the temperature of the Universe drops to ${T\simeq1\,\mathrm{MeV}}$. Below this temperature, they decouple from the thermal bath and flow freely along geodesics of spacetime. Since the neutrinos are still ultrarelativistic when they decouple, they retain a relativistic Fermi-Dirac distribution even though they are no longer in thermal equilibrium. Not being subjected to the Boltzmann exponential suppression, we have far more neutrinos than would otherwise be expected. Although relic neutrinos are very abundant, there is still no direct evidence for their background because it is hard to detect at low energy level, given that they have a very small cross section with matter. There is only indirect evidence for the cosmic neutrino background, mainly through gravitational interactions, for which theoretical predictions are in excellent agreement with observations of the CMB and large-scale structure.

The existence of a neutrino mass has been demonstrated on Earth by neutrino flavor oscillation experiments, which measure the difference between the squares of the mass eigenstates \cite{Super-Kamiokande:1998kpq,T2K:2013ppw}. The experiments can constrain the minimum neutrino mass in two different scenarios (normal or inverted hierarchy) for the ordering of the individual masses. The lowest limit is given by the normal ordering of the mass eigenstates, where the total sum, ${\sum m_\nu>0.06\,\mathrm{eV}}$, is provided by one massive neutrino and two massless neutrinos \cite{Esteban:2018azc,RoyChoudhury:2019hls,Gariazzo:2022ahe}. This in turn implies a lower limit on the current neutrino energy density in standard cosmology, using ${\Omega_\nu h^2\simeq \sum m_\nu/94\,\mathrm{eV}}$, where $\Omega_\nu$ is the neutrino density parameter and $h$ is the current value of the Hubble parameter in units of ${100\,\mathrm{km\,s^{-1}\,Mpc^{-1}}}$. We also know from the Katrin Laboratory's $\beta$-decay experiment that the mass eigenvalues are below the eV scale \cite{KATRIN:2019yun}. But the strongest constraints come from cosmological observations (see, for example, Refs.~\cite{Vagnozzi:2017ovm,Tanseri:2022zfe,Vagnozzi:2018jhn,RoyChoudhury:2018gay,RoyChoudhury:2018vnm}), which provide an upper limit on the current neutrino density. This can be translated into a maximum value for the neutrino mass. Gerstein and Zeldovich originally derived this limit in the 1960s \cite{1966ZhPmR...4..174G}. They showed that the requirement that neutrinos do not overclose the Universe, ${\Omega_\nu h^2<1}$, suggests a cosmological constraint below a hundred eV. This limit was stronger than those of laboratory experiments at the time. Today, it is the Planck 2018 measurements of the temperature and polarization of the CMB \cite{Planck:2018vyg} that provide the most robust constraints: ${\sum m_\nu<0. 26\,\mathrm{eV}}$ at the $95\%$ confidence level (C.L.), for the standard seven-parameter model ${\Lambda\mathrm{CDM}+\sum m_\nu}$ (see neutrinos in cosmology \cite{Workman:2022ynf}).

Alternative cosmological scenarios that extend the parameter space, including modified gravity theories, may relax the neutrino mass bounds \cite{eBOSS:2020yzd,Valentino_2020,Bellomo:2016xhl,Atayde:2023aoj}. For example, fitting a 12-parameter model to Planck and baryon acoustic oscillation (BAO) data increases the limit from ${\sum m_\nu<0.13\,\mathrm{eV}}$ to ${\sum m_\nu<0.52\,\mathrm{eV}}$ (95\% C.L.) \cite{Valentino_2020}. This extensive scenario includes five additional free parameters: two for the parametrization of the dark energy equation of state, one for the running of the spectral index, one for the effective number of relativistic particles, and one unphysical for the amplitude of the dark matter lensing contribution to the CMB power spectra. In the present work, we also investigate whether the existing cosmological upper limit on the neutrino mass can be relaxed. Our proposed model requires only two extra free parameters describing a possible interaction between the neutrino fluid and the dark energy component \cite{SupernovaSearchTeam:1998fmf,SupernovaCosmologyProject:1998vns} given by a scalar field.

We consider a mass-varying neutrino (MaVaN) scenario, where the coupling leads to an effective neutrino mass that depends on the value of the field \cite{Gu:2003er,Fardon:2003eh,Peccei:2004sz,PhysRevD.76.049901,Wetterich:2007kr, Amendola:2007yx,Ichiki:2007ng,Franca:2009xp,Geng:2015haa}. We employ a minimal parametrization where the scalar field depends linearly on the number of $e$-folds \cite{Nunes:2003ff}. It limits the number of additional parameters with respect to the concordance model and alleviates the initial conditions problem \cite{1988ApJ...325L..17P} thanks to the scaling behavior of the field at early times. Such parametrization has been used in the context of testing a coupling between quintessence and the electromagnetic sector, and scalar-field-dark-matter interactions \cite{daFonseca:2022qdf,Barros:2022kpo,daFonseca:2021imp}, but never utilized in the context of neutrino interactions. By testing the model with a particular dataset that combines observations of the CMB, structure growth, and background expansion, we show that the constraint on today's mass is weakened by neutrinos of growing mass \cite{Mota:2008nj,Nunes:2011mw} that receive energy from the quintessence component over cosmic time.

A mechanism that couples the scalar field, as early dark energy, to the neutrinos has been proposed \cite{Sakstein:2019fmf,CarrilloGonzalez:2023lma} to alleviate the Hubble tension, i.e., the discrepancy between the $H_0$ determinations of high and low redshift probes \cite{Planck:2018vyg,Riess:2019cxk,Wong:2019kwg,Riess:2020fzl,Banerjee:2020xcn,Lee:2022cyh}, but it remains to be tested with cosmological observations. Nevertheless, the Hubble tension is not the subject of this paper, since the early dark energy component in our model is insufficient to affect it.

In the next section, we present the MaVaN theory we have chosen to study, together with our specific scalar field parametrization. The phenomenology of the model is analyzed at the background level. In Sec.~\ref{sec:perturbations}, we assess the impact of the interaction on the linear perturbations, as well as the sensitivity of the observables to the coupling. We numerically compute the power spectra of matter, the CMB temperature anisotropies, and the lensing potential with the Einstein-Boltzmann code {\small\texttt{CLASS}} \cite{lesgourgues2011cosmic,class}, which we have modified to compute the theoretical observables of the varying-mass neutrino model. The model is tested against observations in Sec.~\ref{sec:parameter_estimation}. We estimate the cosmological parameters by performing Bayesian inferences on a dataset that combines Planck measurements of the CMB with the detection of baryon acoustic oscillations. We discuss the results in Sec.~\ref{sec:conclusion}, in particular the constraints we obtain on the current neutrino mass sum $\sum m_\nu$ with our chosen dataset in the context of the MaVaN scenario.

\section{Coupling dark energy to neutrinos}\label{sec:model}

Let us consider a flat Universe with vanishing curvature, spatially homogeneous and isotropic, whose expansion is parametrized by the scale factor $a$ associated with the Friedmann-Lema\^itre-Roberson-Walker spacetime metric. Further considering that the expansion is sourced by photons ($\gamma$), baryons (b), cold dark matter (c), neutrinos ($\nu$), and a scalar field dark energy ($\phi$) responsible for the current acceleration, the Friedmann equation reads
\be
\label{Friedmann}
\frac{H^2}{H_0^2}=\Omega_\gamma\,a^{-4}+\left(\Omega_b+\Omega_c\right)a^{-3}+\frac{\rho_\nu(a)+\rho_\phi(a)}{\rho_0}\mathcomma
\ee
where $H\equiv \dot{a}/a$ is the Hubble parameter, the dot denoting derivation with respect to cosmic time $t$. $H_0$ is the current expansion rate, ${\Omega_i=\kappa^2\rho_{i,0}/3H_0^2}$ are the present-day density parameters, $\rho_i$ being the energy density of species $i$ and ${\rho_0={3H_0^2/\kappa^2}}$ the critical density (${\kappa^2\equiv8\pi G=c=\hbar=k_B=1}$ by convention). We use $\rho_\nu$ for the neutrino energy density summed over the mass eigenstates, and we choose a dynamical parametrization for the scalar energy density $\rho_\phi$ in the following \cite{Nunes:2003ff}.

In this study, we want to test a possible interaction between neutrino species and dark energy, in a mass-varying neutrino model where active neutrinos are coupled to the scalar field \cite{Gu:2003er,Fardon:2003eh,Peccei:2004sz,PhysRevD.76.049901,Wetterich:2007kr, Amendola:2007yx,Ichiki:2007ng,Franca:2009xp}. Because to leading order cosmological data are only sensitive to the total neutrino mass \cite{PhysRevD.73.123501,Font-Ribera:2013rwa}, we assume for practical purposes \cite{CORE:2016npo} two massless neutrinos and a massive neutrino nonminimally coupled to the quintessence component. The coupled neutrino has a varying effective mass, which depends on the value of the scalar field and on a dimensionless and constant parameter $\beta$,
\be
\label{varying_mass}
m_\nu(\phi)=m_{\nu,0}\,e^{\beta(\phi-\phi_0)}\mathcomma
\ee
where ${m_{\nu,0}}=\sum m_\nu$ is the current neutrino mass and $\phi_0$ is the current field value. This coupling can be interpreted as a conformal transformation in the Einstein frame \cite{PhysRevLett.64.123}. The massive neutrinos are free falling along the geodesics given by a different metric, $\Tilde{g}^{(\nu)}_{\mu\epsilon}$, which is specific to the neutrino sector only, and is related to the gravitational metric, $g_{\mu\epsilon}$, by a conformal transformation,
\be
\label{conformal_transformation}
\Tilde{g}^{(\nu)}_{\mu\epsilon}=m_\nu^2\left(\phi\right)g_{\mu\epsilon}\mathperiod
\ee
The stress energy tensors of the neutrino fluid and the scalar field are not conserved separately. We have
\begin{align}
    \label{bianchi_neutrino}
    \nabla^\mu T_{\mu\epsilon}^{(\nu)}&=-\beta(\rho_\nu-3p_\nu)\nabla_\epsilon\phi \mathcomma \\
    \label{bianchi_scalar}
    \nabla^\mu T_{\mu\epsilon}^{(\phi)}&=\beta(\rho_\nu-3p_\nu)\nabla_\epsilon\phi \mathcomma
\end{align}
where the subscript $(\nu)$ stands for the stress-energy tensor of the massive neutrino component and $(\phi)$ for that of dark energy, and $p_\nu$ is the pressure in the interacting neutrinos. The time component of Eqs.~\eqref{bianchi_neutrino} and \eqref{bianchi_scalar} give the neutrino and scalar field continuity equations, respectively,
\begin{align}
\label{neutrino_continuity}
\dot{\rho}_\nu+3H\left(\rho_\nu+p_\nu\right)&=\beta\left(\rho_\nu-3p_\nu\right)\dot{\phi}\mathperiod \\
\label{continuity_field}
\dot{\rho}_\phi+3H\left(\rho_\phi+p_\phi\right)&=-\beta\left(\rho_\nu-3p_\nu\right)\dot{\phi}\mathcomma
\end{align}
where $p_\phi$ is the pressure in the field. The extra source terms vanish without interaction, ${\beta=0}$, or if the massive neutrino particles are ultrarelativistic, behaving as traceless radiation.

\begin{figure}[t]
\centering
      \includegraphics[height=0.74\linewidth]{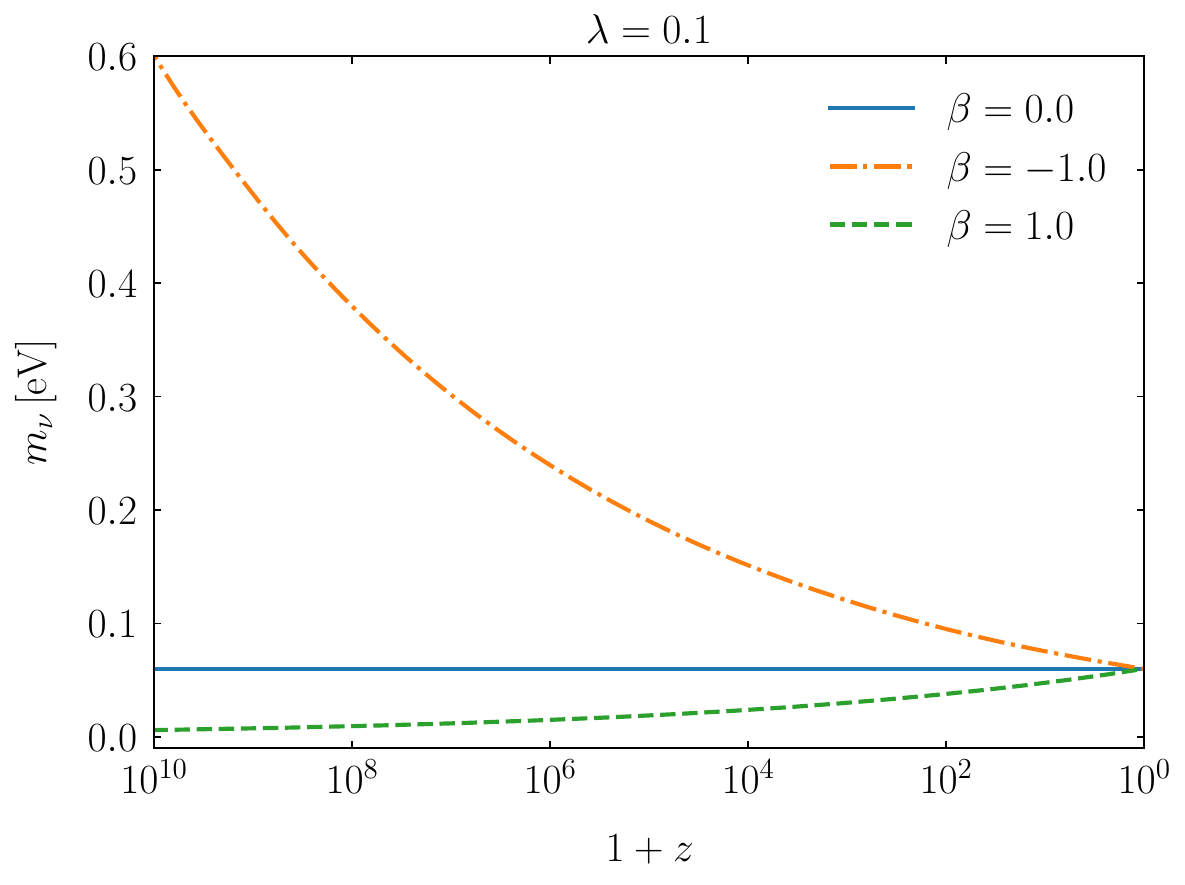}
  \caption{\label{fig:mass} Neutrino mass evolution in growing ($\beta>0$) and shrinking ($\beta<0$) scenarios for ${m_{\nu,0}=0.06\,\mathrm{eV}}$.}
\end{figure}

Regarding the interacting scalar field, we assume that it is homogeneous and canonical with a potential $V(\phi)$ such that ${\rho_\phi=-1/2\,g^{\mu\nu}\partial_\mu\phi\partial_\nu\phi+V}$, and the quintessence pressure is ${p_\phi=\rho_\phi-2\,V}$. Equation~\eqref{continuity_field} can also be written as a modified Klein-Gordon equation, which governs the motion of the scalar field, with an additional term due to the neutrino coupling,
\be
\label{klein_gordon}
\Ddot{\phi}+3H\dot{\phi}+\frac{dV}{d\phi}=-\beta\left(\rho_\nu-3p_\nu\right)\mathperiod
\ee
To test the model with observations, we adopt a known phenomenological parametrization, first proposed in Ref.~\cite{Nunes:2003ff}, where the scalar field depends linearly on the number of $e$-folds, ${N\equiv\ln a}$, throughout the cosmological evolution. We introduce a dimensionless constant $\lambda$ for the slope of the scaling:
\begin{equation}
    \label{lambda}
\phi-\phi_0=\lambda\ln a \mathperiod
\end{equation}
Given the symmetry ${(\phi,\lambda )\mapsto (-\phi ,-\lambda)}$, we will limit the present analysis to ${\lambda>0}$. Also, without loss of generality, we choose to set ${\phi_0=0}$, which implies ${\phi<0}$. Accordingly, we illustrate in Fig.~\ref{fig:mass} the cases ${\beta>0}$ (green dashed line) and ${\beta<0}$ (orange dash-dotted line) which correspond to a growing and to a shrinking neutrino mass, respectively, using Eq.~\eqref{varying_mass}, while the mass is constant for ${\beta=0}$ (blue solid line).

\begin{figure*}[t]
\centering
      \includegraphics[height=0.37\linewidth]{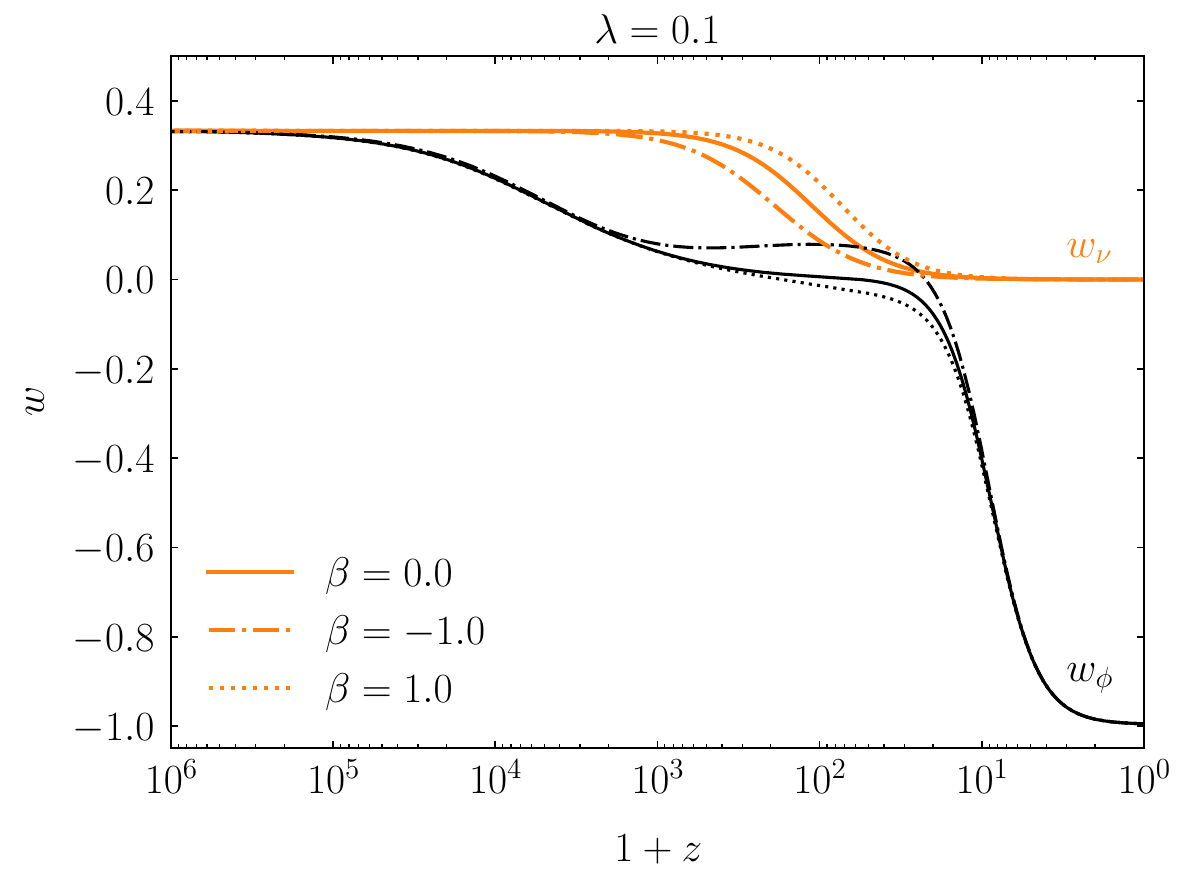}
      \includegraphics[height=0.37\linewidth]{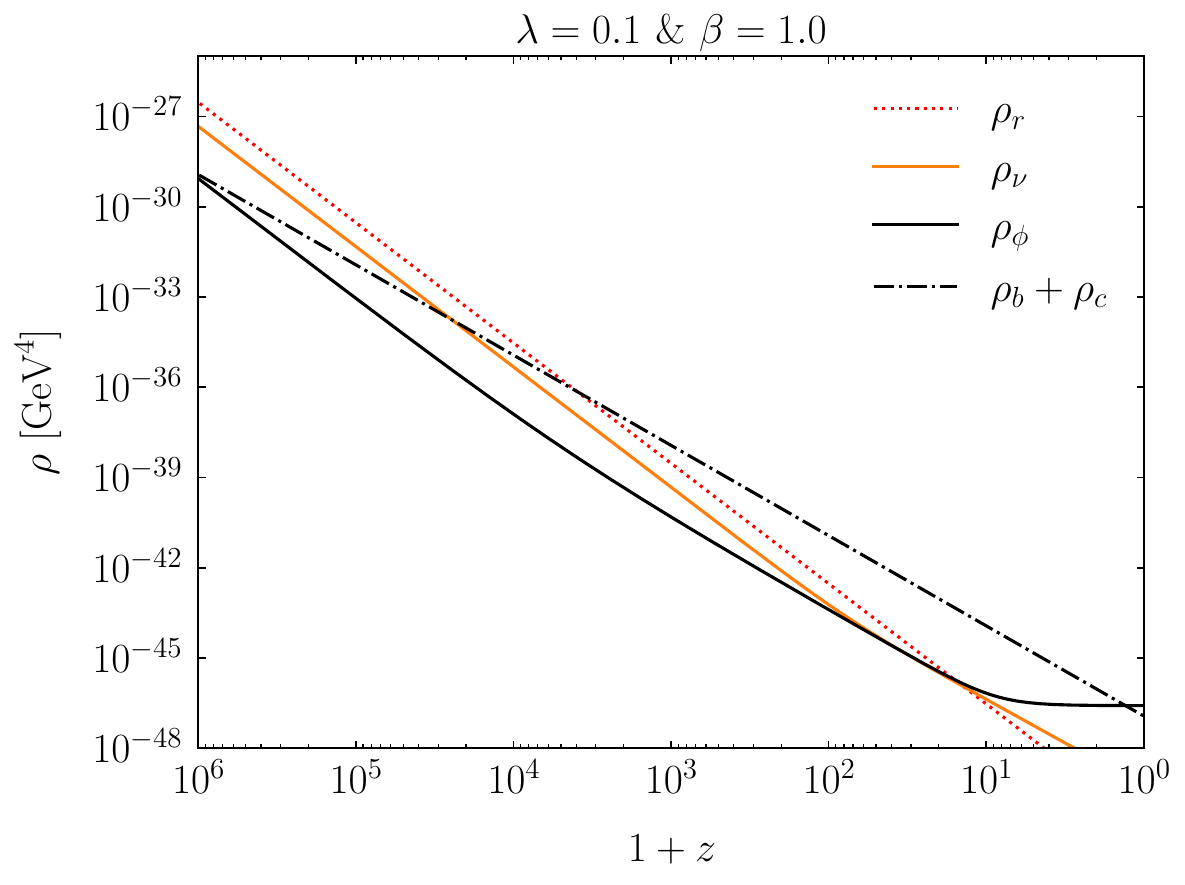}
  \caption{\label{fig:densities_state_neutrino}Left panel: evolution of the equations of state of dark energy ($w_\phi$) and of massive neutrinos ($w_\nu$) for ${m_{\nu,0}=0.06\,\mathrm{eV}}$ and three illustrative values of $\beta$. Right panel: evolution of energy densities for ${m_{\nu,0}=0.06\,\mathrm{eV}}$. The radiation energy density ($\rho_r$) includes photons and ultrarelatistic neutrinos.}
\end{figure*}

This simple approach is a powerful alternative to the popular Chevallier–Polarski–Linder (CPL) parametrization \cite{param_a1,param_a2}, since a large variety of dark energy equation of state evolutions can be captured by just one additional parameter \cite{Nunes:2004eog}, thereby limiting the degeneracies in Bayesian inferences. The background evolution of the dark energy component allows for an early dark energy component at high redshift, something that is not present in the CPL option. Our scalar parametrization manages to represent a vast class of dark energy models that rely on scaling or tracking solution. It does not, however, allow phantom models, where the scalar field equation of state is below $-1$. These are known to provide weaker constraints on the neutrino mass \cite{CORE:2016npo}. It also does not allow an oscillatory evolution of the field around the minimum of an effective potential, such as the model in Ref. \cite{Mota:2008nj}.

An additional advantage is that the scalar field potential can be reconstructed analytically following Refs.~\cite{Nunes:2003ff,daFonseca:2021imp,daFonseca:2022qdf,Barros:2022kpo}. This is done by solving the first-order differential equation \eqref{continuity_field} to find $\rho_\phi$ using the constraint equation \eqref{Friedmann} and noting that ${\dot{\phi}=\lambda H}$ according to Eq.~\eqref{lambda}. The potential happens to be a sum of exponential terms,
\be
\label{potential}
V\left(\phi\right)=A\,e^{-\frac{4}{\lambda}\phi}+B\,e^{-\frac{3}{\lambda}\phi}+C\,e^{\left(-\frac{3}{\lambda}+\beta\right)\phi}+D\,e^{-\lambda\phi}\mathcomma
\ee
where the mass scales are given by the following analytical expressions:
\begin{align}
\label{mass_scales}
A&=\frac{\lambda^2}{4-\lambda^2}\Omega_rH_0^2\mathcomma \\
B&=\frac{3}{2}\frac{\lambda^2}{3-\lambda^2}\left(\Omega_b+\Omega_c\right)H_0^2\mathcomma \\
C&=\frac{3}{2}\frac{\lambda^2+2\beta\lambda}{3-\lambda^2-\beta\lambda}\Omega_\nu H_0^2\mathcomma \\
D&=3\left(1-\frac{\lambda^2}{6}\right)H_0^2\left[1+\frac{4}{\lambda^2-4}\Omega_r \right. \\ \nonumber
& \left. +\frac{3}{\lambda^2-3}\left(\Omega_b+\Omega_c\right)+\frac{3}{\lambda^2-3+\beta\lambda}\Omega_\nu\right] \mathcomma
\end{align}
$\Omega_r$ includes both photons and ultrarelativist neutrinos. As for the density parameter $\Omega_\nu$, it corresponds to those neutrinos that are nonrelativistic, and we can write \cite{lesgourgues2013neutrino}
\be
\label{neutrino_density}
\Omega_\nu h^2=\frac{m_{\nu,0}}{93.14\,\mathrm{eV}}\mathperiod
\ee

We specifically modify the {\small\texttt{CLASS}} code to evolve the scalar field with the potential \eqref{potential} and the equation of motion \eqref{klein_gordon}. The field parametrization is not implemented as such. Independently of the initial conditions, the potential leads to two successive scaling regimes \cite{PhysRevD.61.127301}, where the field equation of state, shown in the left panel of Fig.~\ref{fig:densities_state_neutrino}, first tracks radiation (${w_\phi\sim1/3}$) and then matter (${w_\phi\sim0}$), before being attracted to the late-time acceleration stage (${w_\phi\sim-1}$), always present when ${\lambda^2<2}$. At this point, the energy density of the field freezes, mimicking a cosmological constant at late time, as shown in the right panel.

\begin{figure}[t]
\centering
      \includegraphics[height=0.74\linewidth]{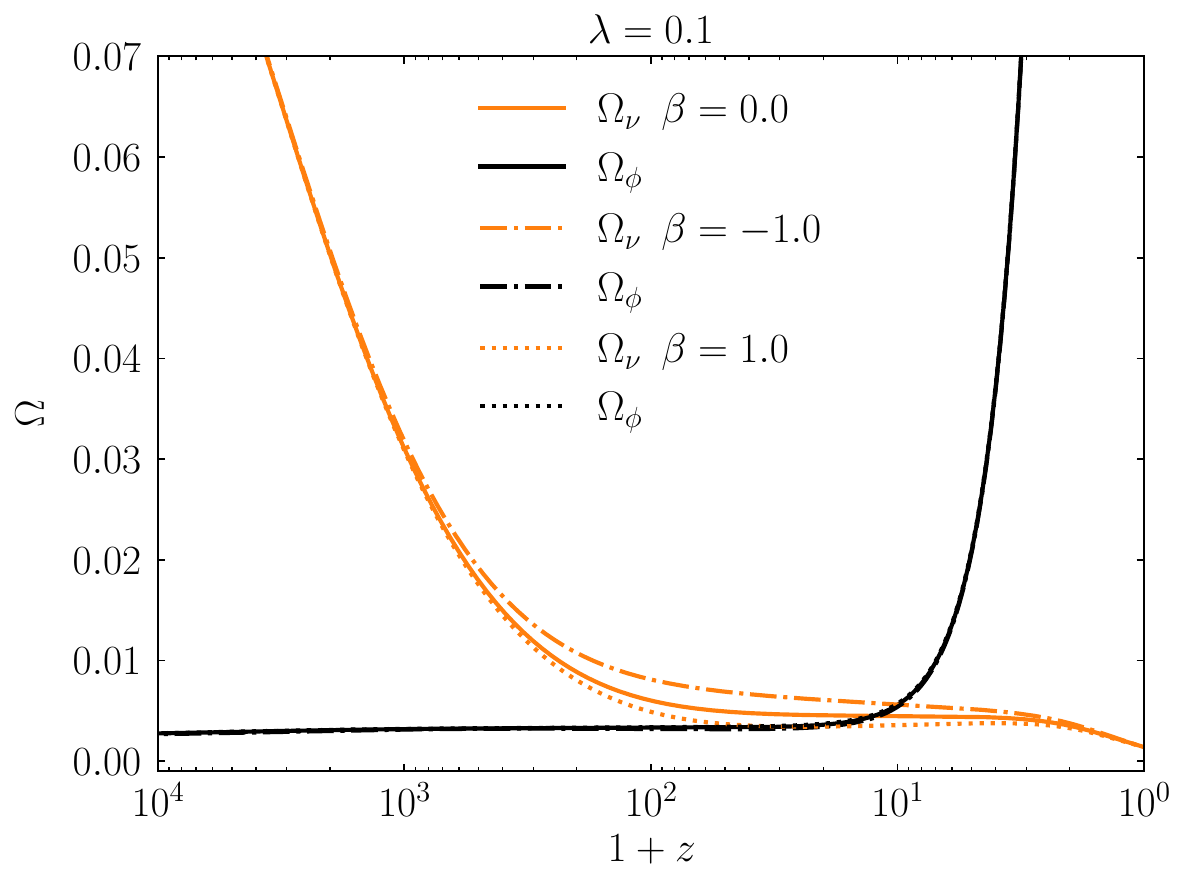}
  \caption{\label{fig:abundances} Evolution of the abundances of massive neutrinos and dark energy energy for ${m_{\nu,0}=0.06\,\mathrm{eV}}$.}
\end{figure}

We can see from Fig.~\ref{fig:densities_state_neutrino} that the coupling with neutrinos changes $w_\phi$ during the matter-dominated era. For growing masses (${\beta>0}$, dotted line), the field equation of state is smaller compared to the uncoupled case (${\beta=0}$, solid line). On the contrary, $w_\phi$ is larger when the energy transfer occurs in the opposite direction, i.e., from neutrinos of shrinking mass (${\beta<0}$, dash-dotted line). Correspondingly, Fig.~\ref{fig:abundances} shows that the nonrelativistic neutrinos which receive energy from the scalar field ($\beta>0$) have lower fractional energy density to reach the same present mass than when they give energy ($\beta<0$).

In the early Universe, neutrinos are held in equilibrium in the primordial plasma by electroweak interactions with charged leptons tightly coupled to photons by electromagnetic processes. As the temperature of the Universe decreases, the interaction rate drops much faster than the Hubble expansion rate. Around ${1\,\mathrm{MeV}}$, at redshift ${z=1/a-1\sim10^{10}}$, neutrinos cease to interact with the cosmological plasma and flow freely along geodesics. Unlike other particles, the neutrinos are hot relics that decouple while still ultrarelativistic. Therefore their unperturbed phase-space-distribution function maintains to a good approximation the Fermi-Dirac shape for an ultrarelativistic fermion in thermal equilibrium,
\be
f_0(q)=\frac{1}{e^q+1}\mathcomma
\ee
neglecting the chemical potential for neutrinos and antineutrinos (i.e., assuming vanishing lepton asymmetry). ${q=ap/T_{\nu,0}}$ is their normalized comoving momentum, $p$ being the physical momentum, and $T_{\nu,0}$ their current temperature. The unperturbed energy density and pressure of the massive neutrino species are thus given by
\begin{align}
\label{neutrino_density}
\rho_\nu&=\frac{1}{\pi^2}\left(\frac{T_{\nu,0}}{a}\right)^4\int_0^\infty f_0(q)\,dq\,q^2\epsilon\mathcomma\\ \label{neutrino_pressure}
p_\nu&=\frac{1}{3\pi^2}\left(\frac{T_{\nu,0}}{a}\right)^4\int_0^\infty f_0(q)\,dq\,\frac{q^4}{\epsilon}\mathcomma
\end{align}
with 
\be
\label{epsilon}
\epsilon^2=q^2+\frac{a^2\,m_\nu^2(\phi)}{T_{\nu,0}^2}\mathcomma
\ee
where $\epsilon$ is the neutrino comoving energy.

The massive neutrinos in the ultrarelativistic regime, that is up to the nonrelativistic transition, behave as standard radiation with their energy density scaling as $a^{-4}$ and ${p_\nu=\rho_\nu/3}$. In the nonrelativistic regime, the coupling term sources the evolution of both the scalar field and the neutrino fluid, the latter behaving as pressureless matter, ${p_\nu=0}$, with its energy density scaling as ${a^{-3}\,e^{\beta\phi}}$. Deep in this regime, the direct integration of Eq.~\eqref{neutrino_continuity} gives
\begin{align}
\label{rho_massive_neutrino}
\rho_\nu&=3H_0^2\Omega_\nu\,e^{-3N+\beta\phi}\mathcomma \\
&=3H_0^2\Omega_\nu\,a^{-3+\beta\lambda} \mathcomma
\end{align}
given that ${\phi\simeq\lambda N\simeq\lambda\ln a}$ in the parametrization.

From the evolution of their equation of state, ${w_\nu=p_\nu/\rho_\nu}$, in the left panel of Fig.~\ref{fig:densities_state_neutrino}, we see that neutrinos with shrinking mass (${\beta<0}$, dash-dotted line) become nonrelativistic earlier than those with growing mass (${\beta>0}$, dotted line), since the latter are lighter in the past and the transition between the two regimes occurs when the average momentum becomes of the order of the mass, ${\langle p\rangle=3.15\,T_\nu=m_\nu}$ \cite{10.3389/fphy.2017.00070}. The redshift $z_\mathrm{nr}$ at which the coupled neutrinos become nonrelativistic is given by \cite{10.3389/fphy.2017.00070}
\begin{equation}
\label{transition}
1+z_\mathrm{nr}\simeq1900\left(\frac{m_{\nu,0}}{\mathrm{eV}}\right)\,e^{\beta\phi_\mathrm{nr}}\mathperiod
\end{equation}
According to our choice for the parametrization, ${\phi_\mathrm{nr}\simeq-\lambda\ln{(1+z_\mathrm{nr})}}$, we get ${e^{\beta\phi_\mathrm{nr}}\simeq(1+z_\mathrm{nr})^{-\beta\lambda}}$, and thus
\begin{equation}
\label{transition_param}
1+z_\mathrm{nr}\simeq\left[1900\left(\frac{m_{\nu,0}}{\mathrm{eV}}\right)\right]^\frac{1}{1+\beta\lambda}\mathcomma
\end{equation}
where the transition between the two regimes depends on the two extra parameters $\beta$ and $\lambda$ and, as usual, on the current neutrino mass $m_{\nu,0}$.

\section{Impact of the coupling on perturbations and observables}\label{sec:perturbations}

\subsection{Perturbation equations}
\label{subsec:perturbations}

We apply the theory of linear perturbations \cite{Lifshitz:1945du} in the synchronous gauge, adopting the usual conventions of Ref.~\cite{Ma:1995ey}. In particular, in this section, the scalars $h$ and $\eta$ represent the metric perturbations, and the energy density fluctuation of the cosmological species $i$ is described by the density contrast ${\delta_i\equiv\delta\rho_i/\Bar{\rho_i}}$. The overbar denotes the background quantities. The perturbed energy density and pressure of a given species $i$ are ${\delta\rho_i\equiv\rho_i(\vec{x},\tau)-\Bar{\rho}_i(\tau)}$ and ${\delta p_i\equiv p_i(\vec{x},\tau)-\Bar{p}_i(\tau)}$, respectively. In this section, the dot denotes derivation with respect to conformal time $d\tau\equiv dt/a$.

\begin{figure*}[t]
\centering
      \includegraphics[height=0.37\linewidth]{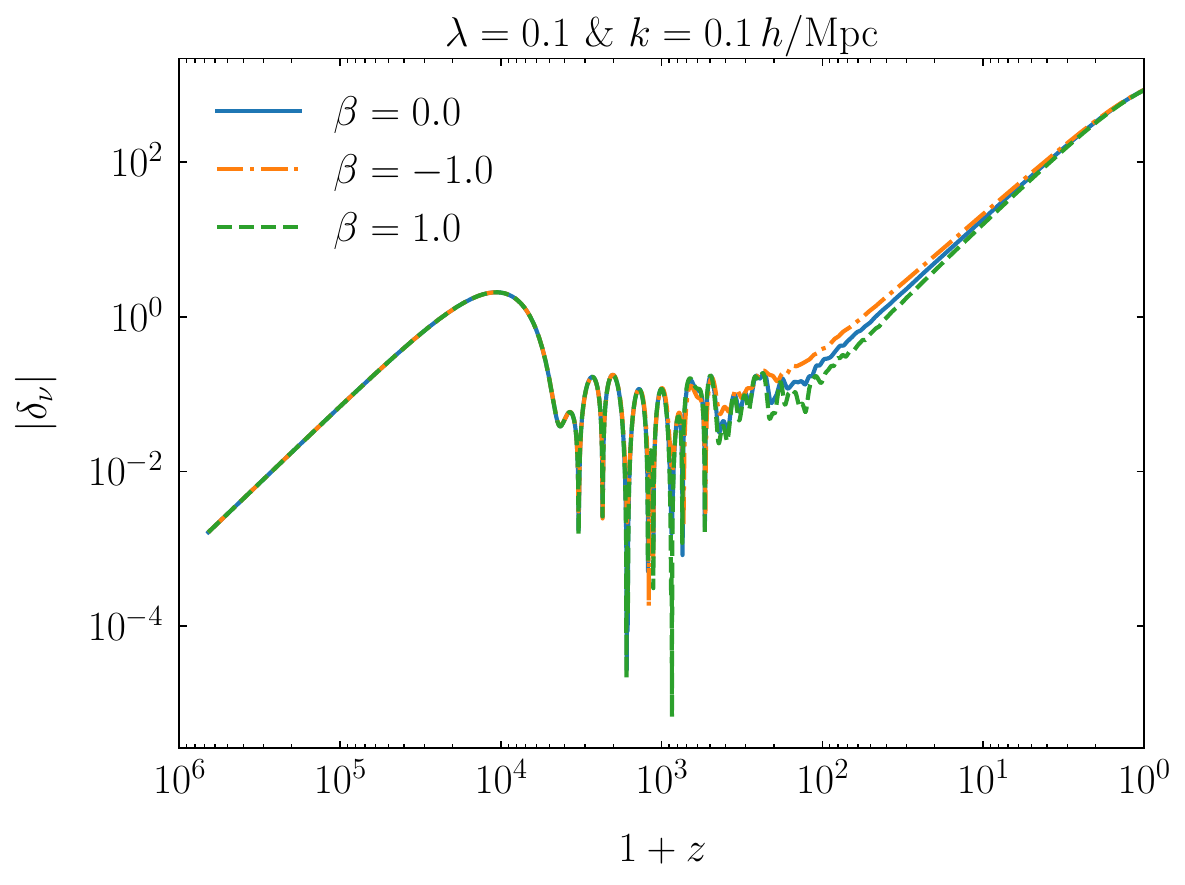}
      \includegraphics[height=0.37\linewidth]{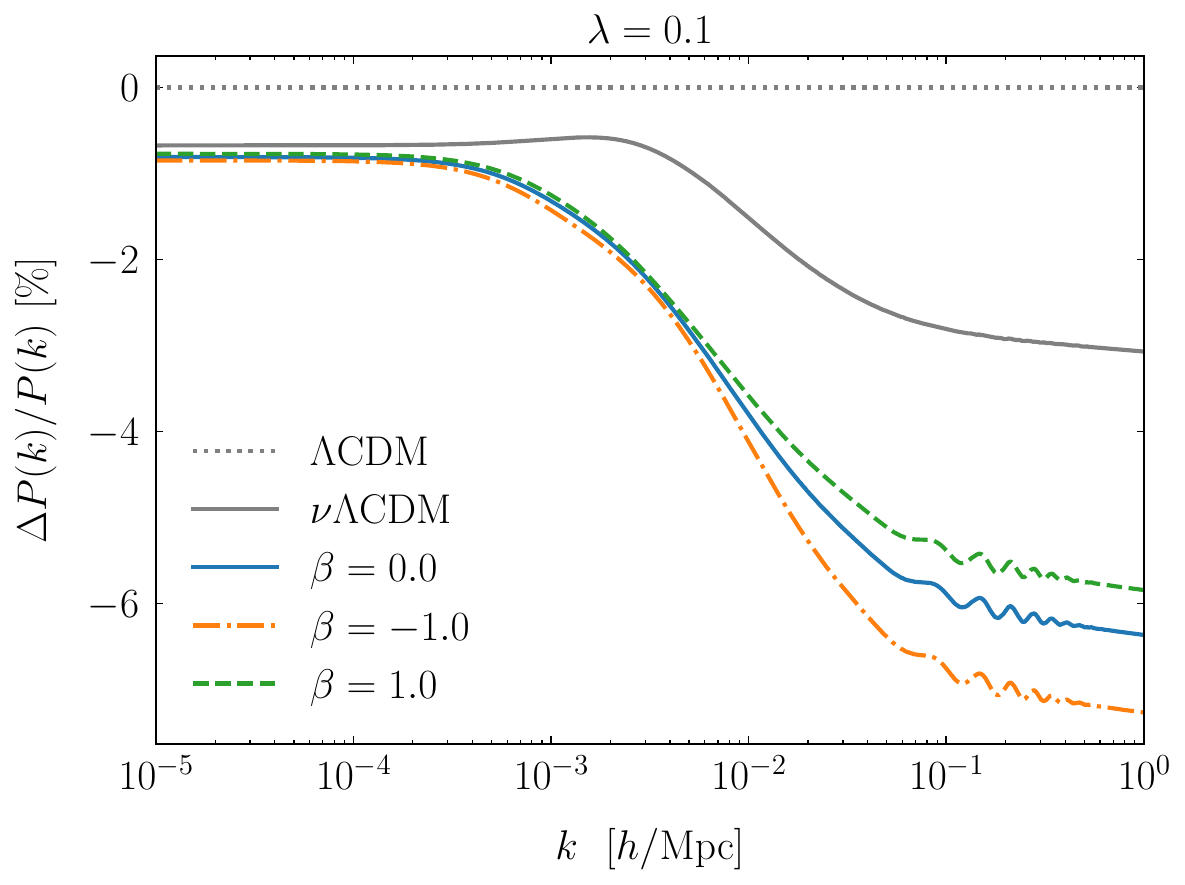}
  \caption{Left panel: evolution of the massive neutrino perturbations at the scale ${k=0.1\,h/\mathrm{Mpc}}$ for ${\lambda=0.1}$ and ${m_{\nu,0}=0.06\,\mathrm{eV}}$. Right panel: suppression of the linear matter power spectrum at $z=0$ with respect to the $\Lambda$CDM model.}
  \label{fig:neutrino_perturbation}
\end{figure*}

The perturbed energy density and pressure of the interacting neutrinos have been derived in previous studies (see, e.g., \cite{Ichiki:2007ng,PhysRevD.76.049901,Franca:2009xp}):
\begin{align}
\label{perturbed_neutrino_density}
\delta\rho_\nu&=4\pi\left(\frac{T_{\nu,0}}{a}\right)^4\int_0^\infty f_0(q)\,dq\,q^2\left(\epsilon\Psi+\beta\delta\phi\frac{m^2_\nu a^2}{\epsilon}\right)\mathcomma\\
\label{perturbed_neutrino_pressure}
\delta p_\nu&=\frac{4\pi}{3}\left(\frac{T_{\nu,0}}{a}\right)^4\int_0^\infty f_0(q)\,dq\,\frac{q^4}{\epsilon}\left(\Psi-\beta\delta\phi\frac{m^2_\nu a^2}{\epsilon^2}\right)\mathcomma
\end{align}
where ${\delta\phi\equiv\phi(\vec{x},t)-\Bar{\phi}(t)}$ denotes the fluctuations of the coupled scalar field, and ${\lvert\Psi\rvert\ll1}$ is the relative perturbation to the neutrino phase-space distribution,
\be
\label{Psi}
\Psi\left(x^i,q,n_j,\tau\right)= \frac{f\left(x^i,q,n_j,\tau\right)}{f_0\left(q\right)}-1\mathcomma
\ee
at first order in the perturbations, $f$ being the perturbed distribution function, and ${n^in_i=1}$. The dipole equation for the neutrino hierarchy is affected by the interaction, and the corresponding system of infinite equations becomes the following in Fourier space:
\begin{align}
    \label{neutrino_hierarchy}
    \dot{\Psi}_0 &=-\frac{qk}{\epsilon}\Psi_1+\frac{\dot{h}}{6}\frac{d\ln f_0}{d\ln q}\mathcomma\\
    \dot{\Psi}_1 &=\frac{qk}{3\epsilon}\left(\Psi_0-2\Psi_2\right)-\frac{qk}{3\epsilon}\beta\delta\phi\frac{a^2m_\nu^2}{q^2}\frac{d\ln f_0}{d\ln q}\mathcomma\\
    \dot{\Psi}_2 &=\frac{qk}{5\epsilon}\left(2\Psi_1-3\Psi_3\right)-\left(\frac{\dot{h}}{15}+\frac{2\dot{\eta}}{5}\right)\frac{d\ln f_0}{d\ln q}\mathcomma\\
    \dot{\Psi}_{\ell\geq3}&=\frac{qk}{\left(2\ell+1\right)\epsilon}\left[\ell\Psi_{\ell-1}-\left(\ell+1\right)\Psi_{\ell+1}\right]\mathcomma 
\end{align}
where $\Psi_\ell$ is the $\ell$th Legendre component of the series expansion in multipole space of the perturbation $\Psi$. We modified the noncold dark matter part of the {\small\texttt{CLASS}} code \cite{Lesgourgues_2011} to evolve the perturbation equations of the MaVaN model. 

In the fluid approximation, on sub-Hubble scales, the Boltzmann hierarchy in the momentum grid is cut at ${l_\mathrm{max}=2}$ and the continuity equation reads
\begin{align}
\label{continuity_perturbation}
\dot{\delta}_\nu=&\,3\left(aH+\beta\dot{\phi}\right)\left(w_\nu-\frac{\delta p_\nu}{\delta\rho_\nu}\right)\delta_\nu \nonumber\\
&-\left(1+w_\nu\right)\left(\theta_\nu+\frac{\dot{h}}{2}\right)
+\beta\left(1-3w_\nu\right)\delta\dot{\phi}\mathcomma
\end{align}
where $\theta_\nu$ is the neutrino flux divergence. Moreover, the Euler equation is
\begin{align}
\label{euler_perturbation}
\dot{\theta}_\nu=&-aH\left(1-3w_\nu\right)\theta_\nu-\frac{\dot{w}_\nu}{1+w_\nu}\theta_\nu+\frac{\delta p_\nu/\delta\rho_\nu}{1+w_\nu}k^2\delta_\nu \nonumber\\
&+\beta\frac{1-3w_\nu}{1+w_\nu}k^2\delta\phi-\beta\left(1-3w_\nu\right)\dot{\phi}\,\theta_\nu-k^2\sigma_\nu\mathcomma
\end{align}
where the neutrino anisotropic stress $\sigma_\nu$ \cite{Ma:1995ey} is not changed by the coupling. We have adjusted the fluid approximation equations of the noncold dark matter in the {\small\texttt{CLASS}} code accordingly.

Deep in the nonrelativistic regime, when ${w_\nu=0}$, the ratio ${q/\epsilon}$ vanishes asymptotically and the pressure perturbations in the neutrino fluid, as well as the shear stress, become negligible with respect to density perturbations. The continuity and Euler equations are analogous to those of the coupled cold dark matter model \cite{Amendola:1999dr,daFonseca:2021imp},
\begin{align}
    \label{NR_regime}
    \dot{\delta}_\nu+\theta_\nu+\frac{\dot{h}}{2}&=\beta\delta\dot{\phi}\mathcomma \\
    \dot{\theta}_\nu+aH\theta_\nu&=\beta\left(k^2\delta\phi-\dot{\phi}\theta_\nu\right)\mathperiod
\end{align}

For the coupled scalar field, the equation of motion of the fluctuations is the following,
\begin{align}
    \label{fluctuations_scalar_field}
    \delta\Ddot\phi+2aH\delta\dot\phi+\left(k^2+a^2\frac{d^2V}{d\phi^2}\right)\delta\phi+\frac{1}{2}\dot{h}\dot{\phi} \nonumber \\
    =-a^2\beta\left(\delta\rho_\nu-3\delta p_\nu\right)\mathperiod
\end{align}
As in the background, we evolve the field perturbations with the potential through the above equation in our version of the {\small\texttt{CLASS}} code.

\subsection{Effects on the matter power spectrum}
\label{subsec:matter_power_spectrum}

There are three main stages of the evolution of the neutrino density contrast affected by the coupling. During the radiation-dominated era, when the neutrinos are decoupled from the thermal bath but still relativistic, their perturbations grow as radiation. Later, the neutrinos become nonrelativistic and cluster in the gravitational potential wells of cold dark matter, which is the dominant cosmological component. However, below their free-streaming scale, they do not cluster like cold dark matter \cite{Lesgourgues:2006nd}. Neutrino free streaming dampens the neutrino fluctuations up to a critical scale depending on the neutrino mass, and giving the oscillatory pattern seen in the left panel of Fig.~\ref{fig:neutrino_perturbation}. The free-streaming wave number of Fourier mode reaches a minimum at the nonrelativistic transition, given by \cite{10.3389/fphy.2017.00070}
\be
\label{free-streaming}
k_\mathrm{fs}\simeq0.018\,\Omega_m^{1/2}\left(\frac{m_{\nu,0}}{\mathrm{eV}}\right)^{1/2}e^{\beta\phi_\mathrm{nr}/2}\,h\,\mathrm{Mpc}^{-1}\mathcomma
\ee
during matter or dark energy domination. Or equivalently, using Eqs.~\eqref{transition} and \eqref{transition_param}, we get
\be
\label{free-streaming_param}
k_\mathrm{fs}\simeq4.1\cdot10^{-4}\,\Omega_m^{1/2}\left(1900\frac{m_{\nu,0}}{\mathrm{eV}}\right)^{\frac{1}{2(1+\beta\lambda)}}\,h\,\mathrm{Mpc}^{-1}\mathcomma
\ee
for our particular scalar field parametrization. Above the free-streaming length, the neutrino fluctuations grow unhindered. For growing neutrino masses (${\beta>0}$, green dashed line) the free-streaming scale in Eq.\eqref{free-streaming_param} is larger and the growth of the fluctuations is delayed with respect to shrinking neutrino masses (${\beta<0}$, orange dash-dotted line).

\begin{figure*}[t]
\centering
      \includegraphics[height=0.37\linewidth]{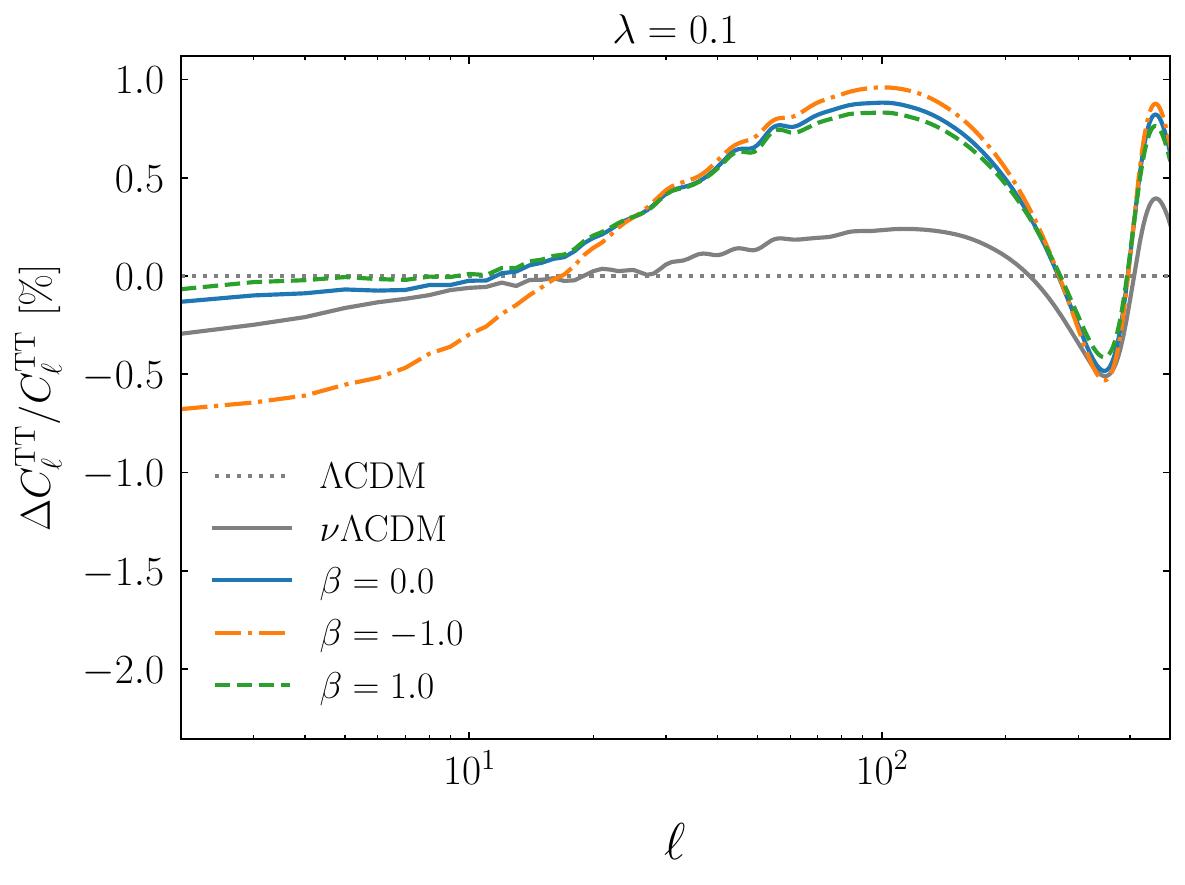}
      \includegraphics[height=0.37\linewidth]{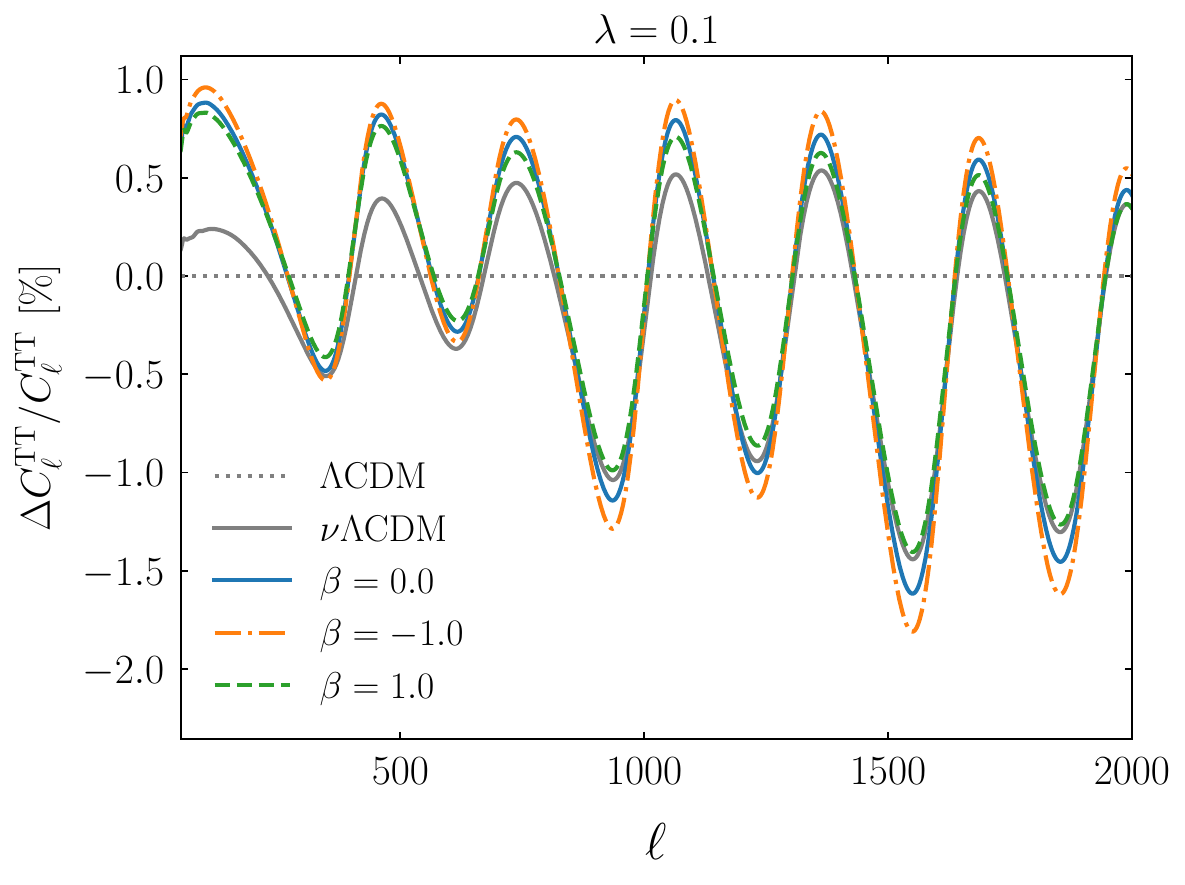}
  \caption{Left panel: comparison of the power spectra of the CMB anisotropies with respect to $\Lambda$CDM, for ${\lambda=0.1}$ and ${m_{\nu,0}=0.06\,\mathrm{eV}}$, on large and intermediate scales. Right panel: same as left, on medium and small scales.}
  \label{fig:CMB}
\end{figure*}

Furthermore, the dependence of the neutrino mass on $\beta$ changes the fraction of matter whose fluctuations do not grow like cold dark matter at a given scale. The neutrinos do not contribute to the creation of potential wells below the free streaming scale, and all structure formation is damped because the gravitational wells are not as deep as they would be in the presence of only nonrelativistic matter.

The effect of the coupling on the linear matter power spectrum, which is proportional to the variance of the density fluctuations (which tells how large they are at a given scale), can be seen on small scales, at large wave numbers $k$. The right panel of Fig.~\ref{fig:neutrino_perturbation} shows the residual plot between the MaVaN scenario and the standard $\Lambda$ cold dark matter (CDM) model with massless neutrinos, normalized to the power spectrum of the latter. At scales smaller than the critical scale ($k\gtrsim10^{-3}$), below which neutrinos do not cluster, the perturbations in the neutrino fluid are completely damped by free streaming and do not contribute to matter perturbations. In this respect, we see with the gray solid line that the ${\nu\Lambda\mathrm{CDM}}$ model\footnote{We use ${\nu\Lambda\mathrm{CDM}}$ to refer to the flat and uncoupled $\Lambda$CDM model with two massless neutrino species and one constant mass neutrino species.} suppresses power at these scales compared to the $\Lambda$CDM model with massless neutrinos.

Moreover, the non-negligible fraction of dark energy itself (${\lambda\neq0}$ and ${\beta=0}$, blue solid line) further reduces the growth of the fluctuations during matter dominance, leading to more power suppression. On the other hand, the matter power spectrum at small scales also depends on how large the neutrino mass was in the past. Growing neutrino masses (${\beta>0}$, green dashed line) reduce the power suppression caused by the scalar field, while shrinking neutrino masses increase the suppression (${\beta<0}$, orange dash-dotted line).

On the contrary, it can be seen that $\beta$ has little influence at large scales ($k\lesssim10^{-3}$), since the neutrinos cluster independently of their mass.

\subsection{Effects on the CMB anisotropies power spectrum}
\label{CMB}

The temperature power spectrum of the unlensed CMB is also affected by the coupling through background and perturbation effects. After their nonrelativistic transition, the massive neutrinos modify the evolution of the cosmological background and the redshift of the matter-dark energy equality, with respect to an uncoupled neutrino universe, affecting the late time integrated Sachs-Wolfe effect. As shown in the left panel of Fig.~\ref{fig:CMB}, at large angular scales (lower multipoles ${\ell\lesssim20}$) reminiscent of the scale-invariant primordial spectrum of the inflationary perturbations, the additional power produced by the time evolution of the gravitational potentials, as dark energy begins to dominate, decreases with negative couplings (orange dash-dotted line).

On the intermediate scales (${20\lesssim\ell\lesssim500}$), dominated by the imprints of the acoustic oscillations in the baryon-photon fluid, it is again the effect of the coupling on the evolution of the gravitational potential that affects the first peak. In particular, $\beta$ alters the time required for the gravitational potential to stabilize at a nearly constant value after photon decoupling. Keeping the present value of the Hubble parameter $h$ unchanged, a higher neutrino mass at the time of recombination (${\beta<0}$, orange dash-dotted line) increases the height of the first peak given by the early integrated Sachs-Wolfe effect compared to the $\Lambda$CDM cases. Conversely, the amplitude decreases for a lower neutrino mass at recombination (${\beta>0}$, green dashed line).

The amplitude of the first peak is also affected by the uncoupled scalar field alone (${\beta=0}$, blue solid line). It gives a non-negligible contribution to early dark energy, which reduces the fractional energy density of matter at the time of decoupling \cite{daFonseca:2022qdf,daFonseca:2021imp}. The amplitude increases with the value of $\lambda$.

Moreover, for a given physical density of cold dark matter and baryons, the angle subtended by the sound horizon at recombination, $\theta_s$, which determines the spacing between the peaks and in particular the position of the first one, is larger for growing neutrino masses ($\beta>0$). The corresponding shift of the position of the CMB peaks towards larger scales can be compensated by the Hubble constant. Indeed, decreasing $H_0$ increases the comoving angular diameter distance from the CMB surface and shifts the peaks towards smaller scales.

At small scales (${\ell\gtrsim500}$), in the right panel of Fig.~\ref{fig:CMB}, of the order of the photon mean free path at the time of recombination, a positive coupling (orange dash-dotted line) has the opposite effect of $\lambda$ (${\beta=0}$, blue solid curve) on the exponential damping of the CMB peak structure.

\subsection{Effects on the CMB lensing potential}
\label{lensing}

\begin{figure}[t]
\centering
      \includegraphics[height=0.96\linewidth]{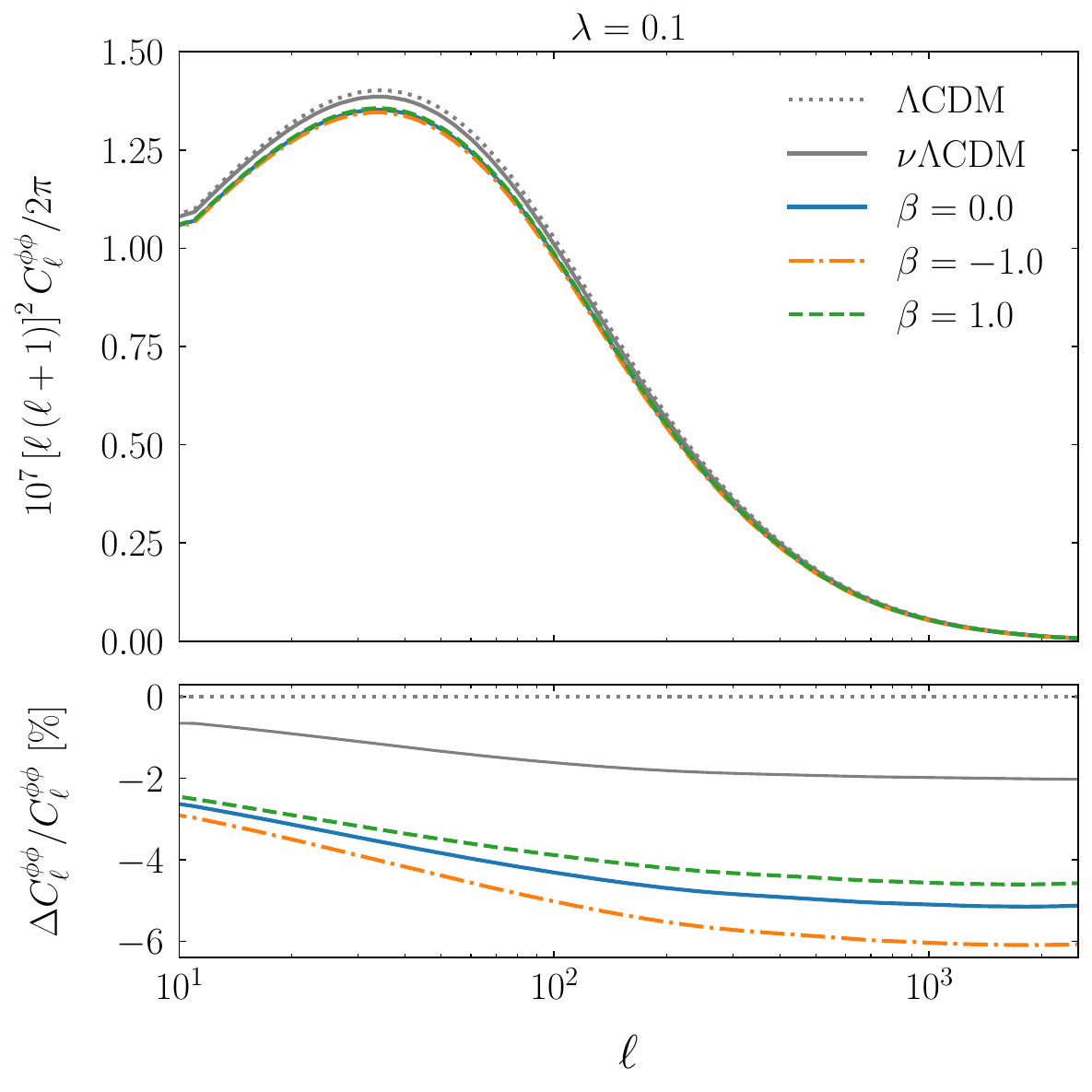}
  \caption{Upper panel: power spectra of the CMB lensing potential of the $\Lambda$CDM and $\nu\Lambda\mathrm{CDM}$ models, and the coupled scenario with $\lambda=0.1$ and $\sum m_\nu=0.06\,\mathrm{eV}$. Lower panel: percentage deviation between models with respect to $\Lambda$CDM.}
  \label{fig:lensing}
\end{figure}

Because the free-streaming neutrinos erase the density perturbations, they affect the CMB light that is distorted by the gravitational lensing caused by the intervening matter distribution between us and the last scattering surface \cite{Lesgourgues:2005yv}. The neutrinos reduce the CMB lensing potential, which is a measure of the integral of the gravitational potentials along the line of sight between the recombination time and the present time. The effect of the weak lensing is to smooth the power spectrum of the CMB temperature anisotropies on small scales. Note in Fig.~\ref{fig:lensing} that since the effect is proportional to the energy density of the neutrinos, it can constrain their mass, whose cosmological evolution is controlled by the two parameters $\lambda$ and $\beta$. For example, if the neutrino mass had been too high in the recent past, we would have had less lensing than we observe. The suppression already caused by the scalar field (${\beta=0}$, blue solid curve) is either enhanced by shrinking neutrino masses ($\beta<0$, orange dash-dotted line) or compensated by growing neutrino masses ($\beta>0$, green dashed line).

It is worth noting that, in contrast to the model-independent parametrization for the neutrino mass variation studied in Ref.~\cite{Franca:2009xp}, we do not find instabilities on large scales in our model \cite{Bjaelde:2007ki}, which would be triggered by large coupling values causing the neutrino perturbations to grow rapidly on the largest scales observable.

\section{Parameter estimation}\label{sec:parameter_estimation}
\subsection{Methodology}\label{subsec:methodology}

We choose to test the model with the 2018 Planck satellite \cite{planck} temperature and polarization measurements of the CMB at the time of last scattering, combined with the CMB lensing potential as a probe of the distribution and evolution of large-scale structure in the late Universe. We also add BAO distance and expansion rate measurements in galaxy clusters  \cite{SDSS:2005xqv,2dFGRS:2005yhx} covering different redshift ranges. They constrain the model at the background level \cite{Percival:2007yw} to help break the geometric degeneracy between $H_0$ and $\sum m_\nu$.

With respect to the Planck observations, the Bayesian inference is performed using the CMB lensing potential power spectrum \cite{Planck:2018lbu}, as well as the cross-correlation likelihoods of the temperature and polarization anisotropies, $TT$, $EE$, and $TE$, respectively \cite{Planck:2019nip}. We reduce the dimensional parameter space by using the lite version of the Planck likelihood on the smaller angular scales, which marginalizes the foreground and instrumental effects, leaving the absolute calibration $A_\mathrm{planck}$ as the only nuisance parameter. As for the background probe, we use a compilation of measurements that have detected the BAO peak at different angular separations with galaxy samples at different redshifts. The three data points are from the Sloan Digital Sky Survey DR7 Main Galaxy Sample \cite{Ross:2014qpa}, the Sloan Digital Sky Survey DR9 release \cite{Ahn_2012}, and the 6dF Galaxy Survey \cite{Beutler_2011}, respectively. We will refer to our dataset as Plk18+BAO in the rest of the document.

Our dedicated version of the {\small\texttt{CLASS}} code is used to compute the observables confronting the actual data, on the basis of the sampled parameters $\{\lambda,\beta,\sum m_\nu\}$ along with the six standard minimal baseline parameters $\{\Omega_bh^2,\Omega_ch^2,\theta_s,A_s,n_s,\tau_\mathrm{reio}\}$. We use $\theta_s$ instead of $h$ because it is the quantity measured in the CMB observations. $A_s$ is the power of the primordial curvature perturbations normalized at the pivot scale ${k=0.05\,\mathrm{Mpc}^{-1}}$, and $n_s$ is the power-law index of the scalar spectrum. $\tau_\mathrm{reio}$ is the optical depth to reionization, which gives the fraction of photons rescattered by the new population of free electrons from the ionization of neutral hydrogen produced by the light of the first stars. The settings in {\small\texttt{CLASS}} are such that the effective number of neutrino families is $N_\mathrm{eff}=3.044$ \cite{Froustey:2020mcq,Bennett:2020zkv,Akita:2020szl}, and the neutrino fractional energy density satisfies Eq.~\eqref{neutrino_density}.

The likelihoods are minimized by the Monte Carlo code of the {\small\texttt{MONTEPYTHON}} parameter estimation package \cite{MP1,BRINCKMANN2019100260}, which samples the parameter space and the posterior probability distributions using a Metropolis-Hastings algorithm with the flat priors specified in Table~\ref{tab:ref}. The size of the prior intervals is sufficient for the corresponding posteriors to fall exponentially within them, except for the parameter $\beta$ as we will discuss later.

\begin{table}[h]
\caption{Flat priors for the sampled model and cosmological parameters.}
 \label{tab:ref}
\centering
\begin{tabular} {l c}
\hline\hline
  Parameter & Prior
  \\
\hline
$\lambda$ & $[0, 1.4]$\\
$\beta$ & $[0, 100]$\\
$\sum m_\nu$& $[0, 2]$\\
\hline
$\Omega_bh^2$& $[0.005, 0.1]$\\
$\Omega_ch^2$& $[0.01, 0.99]$\\
$100\theta{}_{s }$& $[1, 2]$\\
$\ln10^{10}A_{s }$& $[2.7, 4]$\\
$n_s$&  $[0.5, 1.5]$\\
$\tau{}_\mathrm{reio}$& $[0.04, 0.8]$\\
\hline
\end{tabular}
\end{table}

We use the {\small\texttt{GETDIST}} analyzer \cite{Lewis:2019xzd} to obtain the constraints from the Markov chains, and plot the confidence contours and the marginalized posterior distributions. In addition to the sampled parameters, we also infer constraints on derived late-Universe parameters : $H_0$, $\Omega_m$, $\sigma_8$ (which measures the amplitude of matter fluctuation on ${8\,h^{-1}\mathrm{Mpc}}$ comoving scale), and the degeneracy parameter ${S_8\equiv\sigma_8\sqrt{\Omega_m/0.3}}$.

\subsection{Main results}\label{subsec:results}

The results of the likelihood analysis are summarized in Table~\ref{tab:constraints}, which lists the constraints on the cosmological parameters obtained with the Plk18+BAO dataset for both the uncoupled ($\beta=0$) scalar field model and the coupled MaVaN model. They lead to several remarkable conclusions.

\begin{table}[t]
\caption{Cosmological constraints (95\% limits for $\sum m_\nu$, mean and $68\%$ limits for the others) for the uncoupled (${\beta=0}$) and MaVaN ($\beta$ free) models.}
 \label{tab:constraints}
\centering
\begin{tabular} {l c c}
\hline\hline
Parameter & $\beta=0$  & $\hspace{1cm}\beta$ free \\
\hline
$\lambda$ & $< 0.0472$ & $\hspace{1cm}0.059^{+0.029}_{-0.037}$ \\
$\sum m_\nu$& $< 0.127$ & $\hspace{1cm}< 0.724$ \\
\hline
$\Omega_bh^2$& $0.02241\pm 0.00014$ & $\hspace{1cm}0.02242\pm 0.00014$ \\
$\Omega_ch^2$& $0.1199\pm 0.0011$ & $\hspace{1cm}0.1200\pm 0.0011$ \\
$100\theta{}_{s }$& $1.04186\pm 0.00030$ & $\hspace{1cm}1.04179\pm 0.00031$ \\
$\ln10^{10}A_{s }$& $3.050^{+0.014}_{-0.016}$ & $\hspace{1cm}3.049^{+0.014}_{-0.016}$ \\
$n_s$& $0.9667\pm 0.0039$ & $\hspace{1cm}0.9669\pm 0.0039$ \\
$\tau{}_\mathrm{reio}$& $0.0568^{+0.0068}_{-0.0082}$ & $\hspace{1cm}0.0564^{+0.0068}_{-0.0080}$ \\
\hline
$H_0$& $67.59^{+0.69}_{-0.57}$ & $\hspace{1cm}67.61^{+0.65}_{-0.56}$ \\
$\Omega_m$& $0.3127^{+0.0074}_{-0.0091}$ & $\hspace{1cm}0.3176^{+0.0084}_{-0.012}$ \\
$\sigma_8$& $0.813^{+0.012}_{-0.0079}$ & $\hspace{1cm}0.804^{+0.019}_{-0.010}$ \\
$S_8$& $0.830\pm 0.012$ & $\hspace{1cm}0.827\pm 0.015$ \\
\hline
$A_\mathrm{planck}$& $1.0008\pm 0.0025$ & $\hspace{1cm}1.0007\pm 0.0025$ \\
\hline
\end{tabular}
\end{table}

First, in the uncoupled scalar field model (${\beta=0}$), the dynamical dark energy component is constrained to be close to a cosmological constant, ${\lambda< 0.05}$ (68\%\,C.L.) and ${\lambda< 0.09}$ (95\%\,C.L.), with a clear preference for a vanishing scalar field parameter. This is the result of the extreme constraining power of the CMB data on the fraction of early dark energy at the last scattering surface \cite{Gomez-Valent:2021cbe,Gomez-Valent:2022bku}. In our scalar field parametrization, the fraction of dark energy  (during the scaling regime with matter) is given by $\Omega_\phi=\lambda^2/3$ \cite{daFonseca:2021imp}, which implies $\Omega_\phi<0.27\%$ ($95\%\,\mathrm{CL}$) at recombination. It is therefore not surprising that the corresponding upper limit ${\sum m_\nu<0.127\,\mathrm{eV}}$ (95\%\,C.L.) that we find is in agreement with recent literature findings for the concordance model based on comparable datasets. For example, including the latest release of BAO eBOSS data in combination with Planck temperature, polarization, and lensing measurements gives an upper limit of ${0.129\,\mathrm{eV}}$ (95\%\,C.L.) for the uncoupled ${\nu\Lambda\mathrm{CDM}}$ model \cite{eBOSS:2020yzd}.

\begin{figure}[t]
\centering
      \includegraphics[height=1.\linewidth]{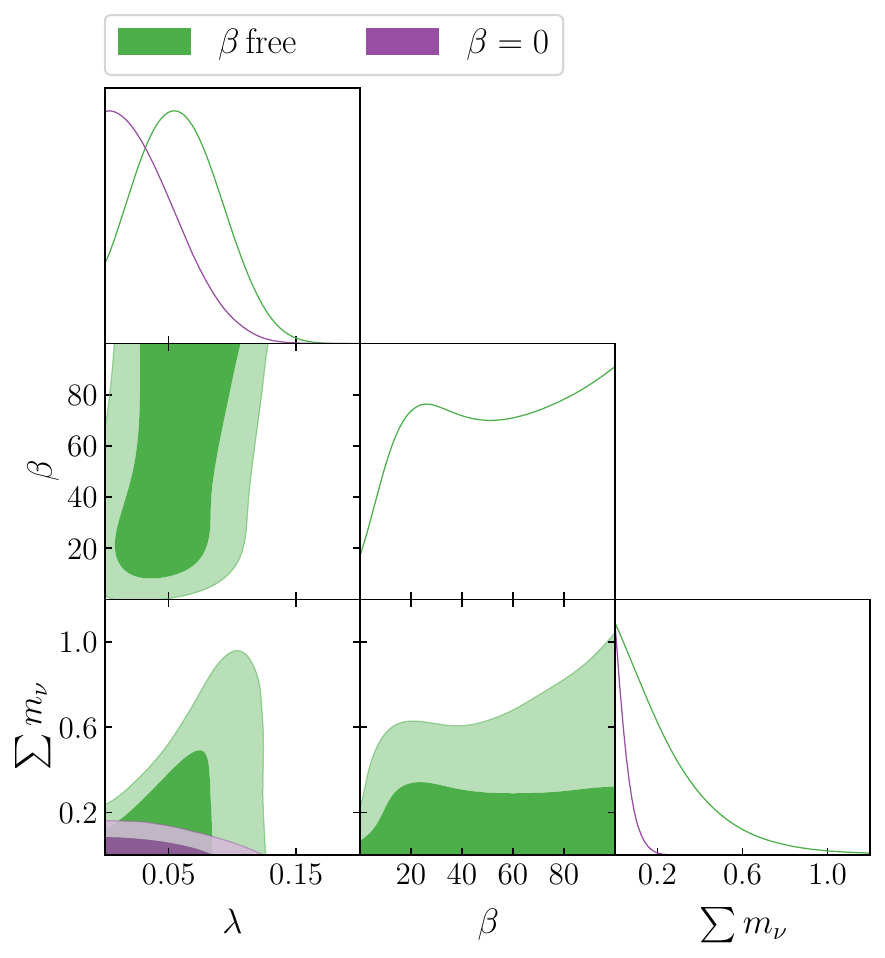}
  \caption{Constraints obtained with the Plk18+BAO dataset on the additional parameters. Probability distributions and 2D
marginalized contours (68\% and 95\% C.L.).}
  \label{fig:beta_free}
\end{figure}

Second, we find that the parameter $\beta$ of the MaVaN model, with our choice of scalar field parametrization, is not constrained by the cosmological observations Plk18+BAO, as shown in Fig.~\ref{fig:beta_free} (and the Appendix for additional contour and probability distribution plots). However, the interaction does relax the upper bound on the current neutrino mass sum, raising it significantly by almost a factor of six to ${\sum m_\nu<0.724\,\mathrm{eV}}$ (95\%\,C.L.). Positive couplings allow for the possibility of heavier neutrinos today. Despite their current higher mass, the influence of neutrinos with growing mass on the cosmological expansion and perturbations over time is tempered by the fact that they were lighter in the past and gradually gained energy from the scalar field. Moreover, for the MaVaN model, the data favor a nonvanishing scalar field parameter, ${\lambda=0.059^{+0.029}_{-0.037}}$ (68\%\,C.L.), at the one-sigma level.

Third, with respect to the standard cosmological parameters, the main differences between the MaVaN model and the uncoupled one are in the matter fluctuation amplitude, $\sigma_8$, and the matter density parameter, $\Omega_m$. Including the coupling parameter degrades the error bars while virtually preserving the central values, which are consistent with the best fit of the base-$\Lambda$CDM cosmology predicted by the 2018 Planck data \cite{Planck:2018vyg}. This is particularly the case for the derived parameter $\sigma_8$, whose precision decreases by about $50\%$, from $0.813^{+0.012}_{-0.0079}$ to $0.804^{+0.019}_{-0.010}$ (68\%\,C.L.), mainly towards lower values, which could tend to reduce the tension with late cosmological probes \cite{DES:2021bvc,Kilo-DegreeSurvey:2023gfr,DES:2021vln}. However, the confidence interval of $\Omega_m$ increases by $22\%$ in the opposite direction due to the added fractional energy density of heavier neutrinos, from $0.3126^{+0.0074}_{-0.0090}$ to $0.3176^{+0.0084}_{-0.012}$ (68\%\,C.L.). As shown in Appendix \ref{appendix}, the posterior distribution for the degeneracy parameter $S_8$ is only slightly changed by the coupling parameter. As for $H_0$, its posterior distribution is hardly modified by the interaction, confirming the inability of the model to resolve the Hubble tension. The fractional dark energy in the early times is far less than the minimum $10\%$ required by neutrino-assisted early dark energy models to improve the $H_0$ tension \cite{Poulin:2018cxd,deSouza:2023sqp}.

\begin{figure*}[t]
\centering
      \includegraphics[height=0.8\linewidth]{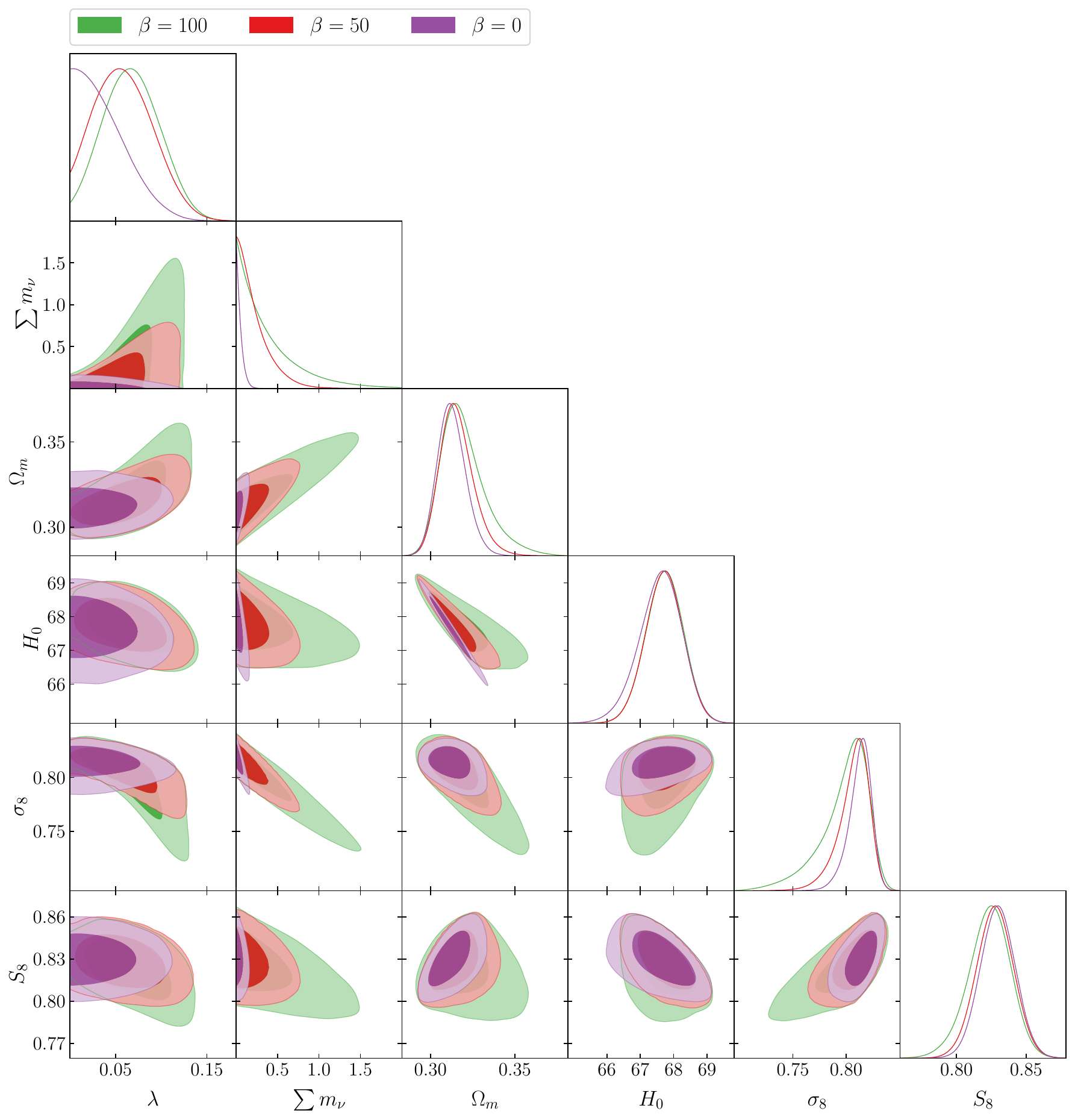}
  \caption{\label{fig:neutrino_comparison}Constraints obtained for fixed coupling parameters. Probability distributions and 2D
marginalized contours (68\% and 95\% C.L.).}
\end{figure*}

Finally, to assess the influence of the coupling on the posterior $\sum m_\nu$, we repeat the statistical analysis fixing different values for $\beta$. The results are summarized in Table~\ref{tab:constraints_beta_fixed} and Fig.~\ref{fig:neutrino_comparison}. For a mass-shrinking neutrino scenario (${\beta=-1}$), the constraint on today's mass is tighter (${\sum m_\nu< 0.102\,\mathrm{eV}}$) than without interaction. Large masses in the past are disfavored by the cosmological data. On the contrary, the limit on today's neutrino masses relaxes as the value of the positive coupling increases. The highest limit is for $\beta=100$, i.e., for the edge of the prior, with $\sum m_\nu< 1.130\,\mathrm{eV}$ already exceeding the limit obtained by local experiments \cite{KATRIN:2019yun}. Moreover, there seems to be a plateau around $\beta=20$ where the present mass stabilizes in the ($\beta,\sum m_\nu$) plane of the parameter space (see also Fig.~\ref{fig:beta_free}). In addition, as seen earlier, the energy transferred by the scalar field causes the model to deviate slightly from a constant dark energy component. Dark energy moves away from a cosmological constant as the strength of the coupling increases. However, the value of the scalar field parameter is tightly constrained by the CMB data, which does not allow $\lambda$ to be too large, while it could theoretically be as large as $\sqrt{2}$.

\begin{table}
    \centering
\caption{Constraints on present neutrino mass in eV and on the dark energy parameter $\lambda$, for different coupling values $\beta$.}
\label{tab:constraints_beta_fixed}
    \begin{tabular}{ccc}
    \hline
    \hline
         $\beta$&  $\sum m_\nu\,(95\%\,\mathrm{C.L.})$ & $\lambda\,(68\%\,\mathrm{C.L.})$\\
         \hline
         -1&  $< 0.102$& $< 0.0497$\\
         0&  $< 0.127$& $< 0.0472$\\
 1& $< 0.147                   $&$< 0.0493                  $\\
         5&  $< 0.355$& $0.046^{+0.016}_{-0.043}$\\
         10& $< 0.580$&$0.050^{+0.022}_{-0.038}$\\
 20& $< 0.542$&$0.052^{+0.024}_{-0.038}$\\
 30& $< 0.513                   $&$0.053^{+0.025}_{-0.039}   $\\
         50&  $< 0.598$& $0.059^{+0.028}_{-0.037}$\\
 75& $< 0.822$&$0.064^{+0.031}_{-0.035}   $\\
         100&  $< 1.130$& $0.067\pm 0.030$\\
         \hline
 free& $< 0.724$&$0.059^{+0.029}_{-0.037}$\\
 \hline
    \end{tabular}
\end{table}

We note in Fig.~\ref{fig:neutrino_comparison} that the $\sum m_\nu$ positive and negative correlations with $\Omega_m$ and $\sigma_8$, respectively, are preserved by the coupling. The errors on $S_8$ are degraded to lower values as the strength of the interaction increases. It is also worth noting that the degeneracy in the ($\sum m_\nu,H_0$) plane, along which the position of the first CMB peak is held unchanged, decreases significantly with increasing value of the parameter $\beta$, weakening the otherwise strong correlation between the neutrino mass and the Hubble constant. This is probably due to the fact that the posterior for $\lambda$ becomes larger with increasing $\beta$. The corresponding non-negligible fraction of dark energy at the time of recombination reduces the sound horizon at decoupling, thus competing with the geometric degeneracy $\sum m_\nu-H_0$.

\section{Discussion} \label{sec:conclusion}

In this study, we investigated a model of mass-varying neutrinos in which the mass depends on the value of a scalar field representing the dark energy component. The originality of our work lies in the choice of the quintessence parametrization, which limits the number of free parameters with respect to other approaches such as the CPL parametrization or arbitrary choices of scalar field potentials. While the parametrization is purely phenomenological, it leads to the analytical reconstruction of the potential as a sum of exponential terms, which conveniently endows dark energy with scaling properties. We have introduced two additional free parameters with respect to $\Lambda$CDM, $\beta$ for the coupling strength between the two sectors in Eq.~\eqref{varying_mass} and $\lambda$ for the linear evolution of the field in Eq.~\eqref{lambda}. 

We confirmed the conjecture that a growing neutrino mass scenario relaxes the upper bound on the current neutrino mass derived in the framework of the baseline $\nu\Lambda$CDM model. This is a result that we expect can be replicated for any quintessence model that manifests scaling or tracking behavior. Our goal was to complement previous work, such as the study in Ref.~\cite{PhysRevD.76.049901}. In the latter, one of the models considered is a mass-varying neutrino theory with an identical coupling to the scalar field, which is a classical choice made throughout the literature, and a different potential in the form of a single exponential. In contrast to our case, their quintessence model shows no tracking behavior and allows only shrinking mass scenarios. The authors also found that astronomical observations do not provide strong constraints on the coupling parameter. To constrain the coupling, they fixed the neutrino mass to values higher than ${0.1\,\mathrm{eV}}$, assuming that such a significant mass could be independently confirmed. Alternately, in our likelihood analysis, we set different values of the coupling to constrain the current neutrino mass.

Following the analysis of the model at background level, we evaluated the sensitivity of several observables (matter and CMB power spectra, and CMB lensing potential) to the coupling, using a version of the Boltzmann code {\small\texttt{CLASS}} that we adapted to the considered model. We found that growing neutrino masses lead to less matter power suppression than predicted by the presence of a noninteracting scalar field. The coupling also affects the shape of the CMB power spectrum at different scales, in particular through the integrated Sachs-Wolfe effect, in parallel with the effect of the scalar field parameter itself. The CMB lensing potential is sensitive to the interaction. Growing neutrino masses can compensate for the reduction in lensing potential caused by the quintessence fluid. It is therefore theoretically possible to obtain constraints on a putative interaction between the neutrino sector and a dynamical dark energy component.

We performed Bayesian inferences to estimate the parameters of the model using observations of CMB temperature and polarization anisotropies and lensing from the Planck satellite, together with BAO measurements. This dataset allows us to combine cosmological probes covering the early Universe at the time of the last scattering, the formation of large-scale structures in the late universe, and the cosmic expansion history. The Markov chains were generated by the {\small\texttt{MONTEPYTHON}} package using a Metropolis-Hastings algorithm to explore the parameter space, which contains nine free parameters, $\{\lambda,\beta,\sum m_\nu,\Omega_bh^2,\Omega_ch^2,\theta_s,A_s,n_s,\tau_\mathrm{reio}\}$, plus one nuisance parameter $A_\mathrm{planck}$. We also derived constraints on four late Universe parameters: $H_0$, $\sigma_8$, $\Omega_m$ and $S_8$.

The main conclusion is that the model relaxes the upper bound on the neutrino mass, from ${\sum m_\nu<0.13\,\mathrm{eV}}$ in $\nu\Lambda$CDM to ${\sum m_\nu<0.72\,\mathrm{eV}}$ in our model (95\%\,C.L.). The dynamical property of dark energy is enhanced because the scalar field looses energy to the neutrino, whose mass grows with time. With a non-vanishing scalar field parameter, ${\lambda=0.059^{+0.028}_{-0.036}}$ (68\%\,C.L.), the data disfavors a static field. This is in contrast with previous work where CMB data constraining a scalar field interacting with dark matter model favors $\Lambda$CDM \cite{daFonseca:2021imp}. The coupling $\beta$ is, however, poorly constrained by the observations of the selected dataset. As for the cosmological parameters of the standard model, it is mostly the confidence intervals for $\sigma_8$ and $\Omega_m$ that are broadened by the interaction, but in a way that weakly affects the posterior distribution of the degeneracy parameter $S_8$. The posterior distribution of $H_0$ is also almost unchanged, as expected. The scalar field parameter is extremely constrained and the fraction of dark energy at the last scattering is too small to make any impact in resolving the Hubble tension.

As regards forthcoming work, it would be worthwhile to complement the CMB constraints that the Planck data place on the parameters of our model with precise measurements of the anisotropies at larger multipoles ($l\gtrsim3000$). These small angular scales measured with sufficient accuracy can reveal coupling signatures, especially when the strength of the interaction is small \cite{Brax:2023rrf}. For example, an indication of nonvanishing coupling between neutrinos and dark matter at the one-sigma level has been found using the Atacama Cosmology Telescope's high-multipole observations of CMB temperature and polarization anisotropies \cite{Brax:2023tvn}. In addition, alternative CMB data on the lensing power spectrum could also be used to further constrain the neutrino-scalar field interaction scenario affecting structure growth \cite{ACT:2023dou}. As for the late-time Universe probes of large-scale structure, it would be adequate to use the Kilo-Degree Survey (KiDS) weak lensing observations \cite{Hildebrandt:2016iqg} to test the MaVaN model, like in the coupled dark matter case \cite{daFonseca:2021imp}.

In the future, combinations of CMB experiments (such as Simons Observatory \cite{Ade_2019}, CMB-S4 \cite{abazajian2016cmbs4}, or LiteBird \cite{2019JLTP..194..443H}) with large-scale structure experiments [for example Dark Energy Spectroscopic Instrument (DESI) \cite{desicollaboration2016desi}, Euclid \cite{laureijs2011euclid}, Large Synoptic Survey Telescope (LSST) \cite{lsstsciencecollaboration2009lsst}, or Square Kilometre Array (SKA) \cite{SKA:2018ckk}] are expected to significantly improve the robustness of the cosmological neutrino mass measurements \cite{Brinckmann_2019,Archidiacono_2017}, possibly allowing us to discriminate between models of interacting dark energy.

\acknowledgments

The authors would like to thank C. van de Bruck and  D. F. Mota for the fruitful discussions. This work is supported by the Fundação para a Ciência e a Tecnologia (FCT) through the research Grants No. UIDB/04434/2020 (DOI: 10.54499/UIDB/04434/2020) and UIDP/04434/2020 (DOI: 10.54499/UIDP/04434/2020), and the BEYLA Project No. PTDC/FIS-AST/0054/2021. V.d.F. acknowledges FCT support through fellowship 2022.14431.BD.

\appendix
\section*{Appendix: Parameter constraints for the MaVaN and the uncoupled scalar field models}\label{appendix}
\begin{figure*}[t]
\centering
      \includegraphics[height=1\linewidth]{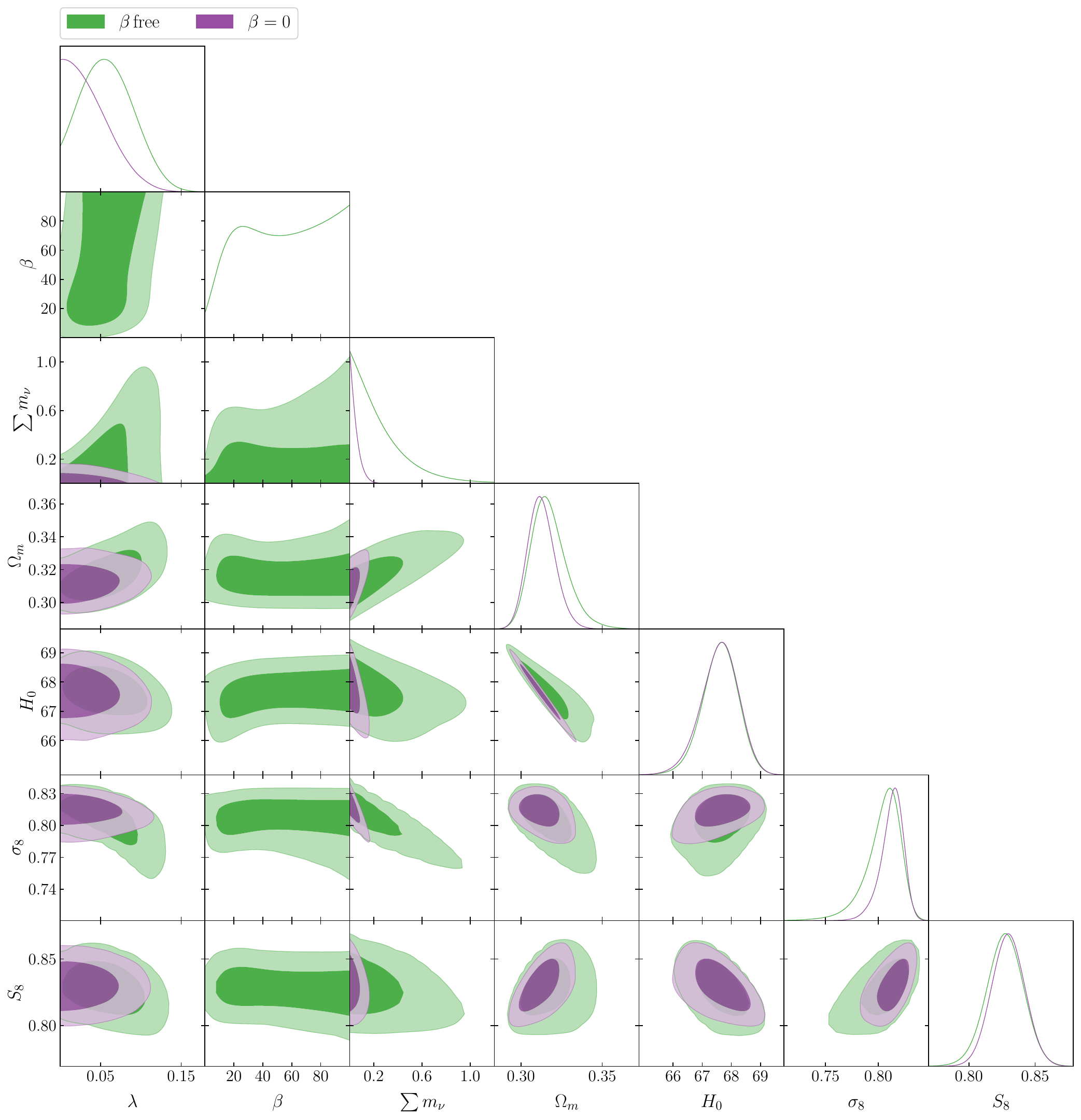}
  \caption{\label{fig:beta_free_all}Probability distributions and 2D marginalized contours (68\% and 95\% C.L.) obtained with the Plk18+BAO dataset.}
\end{figure*}

In this appendix we provide the triangular plots (Fig.~\ref{fig:beta_free_all}) of the analysis made with the Plk18+BAO data for the MaVaN model ($\beta$ free) and the uncoupled case ($\beta=0$), i.e., no interaction between the quintessence component and the neutrino fluid.
\clearpage
\bibliography{bib.bib}
\end{document}